\documentclass[preprint,11pt]{elsarticle}

\usepackage{graphicx}
\usepackage[utf8]{inputenc}
\usepackage[english]{babel}
\usepackage{lineno}
\usepackage{hyperref}
\usepackage{tikzsymbols} 
\usepackage[utf8]{inputenc}
\usepackage[T1]{fontenc}
\usepackage{lmodern}
\usepackage{amssymb}
\begin{document}

\bibliographystyle{elsarticle-num}
\journal{Nucl. Instr. Meth. A}

\title{\bf Nanodiamond photocathodes for MPGD-based single photon detectors}

\begin{frontmatter}
 
\author[b]{F.~M.~Brunbauer} 
\author[a]{C. Chatterjee}
\author[c,d]{G. Cicala}
\author[a,e]{D. D'Ago}
\author[a]{S. Dalla Torre}
\author[d]{M. S. Leone}
\author[a]{S. Levorato}
\author[c,f]{T. Ligonzo}
\author[b]{M.~Lisowska}
\author[a]{R. Rai\corref{cor1}}
\cortext[cor1]{Corresponding author}
\ead{rai.richabhu@gmail.com, richa.rai@cern.ch}
\author[a]{F. Tessarotto}
\author[a,f]{Triloki}
\author[c]{A. Valentini}
\author[g]{L. Velardi}

\affiliation[a]{organization={INFN Trieste},
         country={Italy}}
         
\affiliation[b]{organization={CERN},
            country={Geneva, Switzerland}}   

 \affiliation[c]{organization={INFN Bari},
          country={Italy}}
 \affiliation[d]{organization={CNR-ISTP Bari},
          country={Italy}}
 \affiliation[e]{organization={Dipartimento di Fisica, Universit\`a di Trieste},
        country={Italy}}
\affiliation[f]{organization={Dipartimento Interateneo di Fisica Michelangelo Merlin Universit\`a di Bari},
 country={Italy}}
\affiliation[g]{organization={CNR-IMM Lecce},
            country={Italy}}

\begin{abstract}
This study investigates the suitability of Hydrogenated NanoDiamond (HND) materials as an alternative for CsI in MPGD-based photon detectors. The research focuses on characterizing HND photocathodes coupled with THGEM +Micromegas-based detectors. The HND grains were prepared via hydrogenation and stored in water for more than two years. They were then coated on PCB discs or THGEMs using a pulsed spray technique.  The resulting quantum efficiency (QE) values ($\sim4\%$ at 122 nm)  were found to be within a factor of 10 of the best freshly hydrogenated samples reported in the literature ( $\sim40\%$ at 120 nm). The robustness of reflective HND photocathodes against ion bombardment was measured to be about 10 times larger than the corresponding CsI one after the same charge accumulation. Furthermore, THGEM characterization indicates minimal alteration in response after HND coatings.  These results suggest that HND holds potential as a more robust photocathode for gaseous detectors, offering improved performance in single-photon detection applications.
\end{abstract}


\end{frontmatter}


\section{Introduction}
\label{sec:sample1}
Gaseous Ring Imaging Cherenkov (RICH) detectors are the natural choice to perform hadron identification at high momenta; they require efficient and accurate detection of single Cherenkov photons over large surfaces. MPGDs are used as detectors of single photons, sensitive in the vacuum ultraviolet (VUV) domain~\cite{Silvia}. Presently, CsI is the only photoconverter adequate for gaseous detectors due to its wide wavelength sensitivity (cut-off wavelength  $\sim$ 210 nm) and high QE in VUV ranges, as well as possibility to be deposited over large surfaces. However, the hygroscopic nature of CsI and the limited resistance against ion/photon bombardments are severe limits in terms of QE preservation~\cite{Singh,Hoed}. The quest to overcome these limits and develop a more robust photocathode, capable of coping with challenging conditions motivated the study reported in this article. HND particles have emerged as a potential alternative material with interesting characteristics.

The high QE of CsI photocathodes is attributed to its low electron affinity (0.1 eV) and wide band gap (6.2 eV)~\cite{afinity}. NanoDiamond (ND) particles share similar characteristics with a comparable band gap (5.5 eV) and low electron affinity ranging from 0.35 to 0.50 eV~\cite{Velardi2}. Additionally, ND exhibits chemical inertness and good radiation hardness. The process of ND hydrogenation further reduces the electron affinity to a negative value, enabling efficient escape of generated photoelectrons without encountering an energy barrier at the surface as discussed in \cite{Amos}. A novel ND hydrogenation and coating procedure has been developed at INFN Bari and has provided high and stable QE: 22\% at 146 nm \cite{Velardi,Valentini}, only a factor two less than what Rabus et al. reported for CsI reflective photocathodes (QE of 45\% to 40\% )\cite{Rabus} at the same wavelength. These findings suggest that HND holds promise as a viable alternative to CsI, offering comparable QE values and potentially improved robustness in gaseous detectors applications.  

Our continuing research focuses on the functionality of HND photocathodes coupled with THGEM+Micromegas-based detectors. The study includes the characterization of THGEM coated with HND layers in the single photon detection mode, along with determining the robustness of its photoconverting properties against the bombardment by ions from the multiplication process in the gaseous detector. In addition, the performance of the ND photocathode in $Ar : CH_{4}$ and $Ar:CO_{2}$ gas mixtures is thoroughly investigated. First results on the variation of HND QE in different $Ar-CH_{4}$ mixtures were discussed in \cite{Brunbauer1}. New results of these studies are reported in this article.

\begin{figure}[h]
    \centering
        \includegraphics[clip, trim=0.0cm 0.0cm 0.3cm 0.1cm, width=1.0\textwidth]{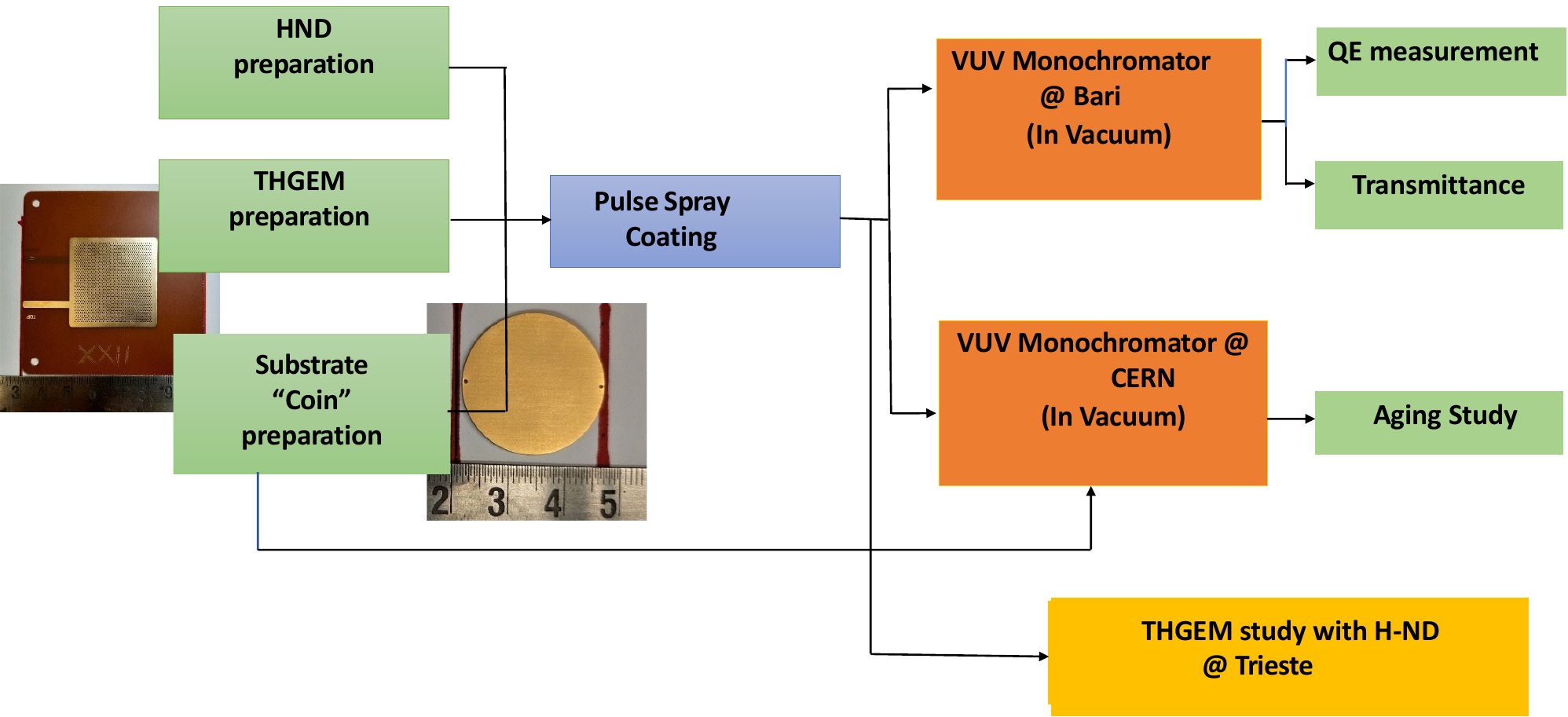}
    \caption{Scheme of the R\&D protocol followed for ND characterization.}
    \label{fig:RD}
\end{figure}
\section{R \& D Methods}

The R \& D protocol implemented for the characterization of ND involves a systematic approach for their successful integration with MPGDs. The scheme followed in the current work is outlined in Fig. \ref{fig:RD}

\begin{figure}[h]
    \centering
        \includegraphics[clip, trim=0.00cm 0.0cm 0.0cm 0.0cm, width=0.90\textwidth]{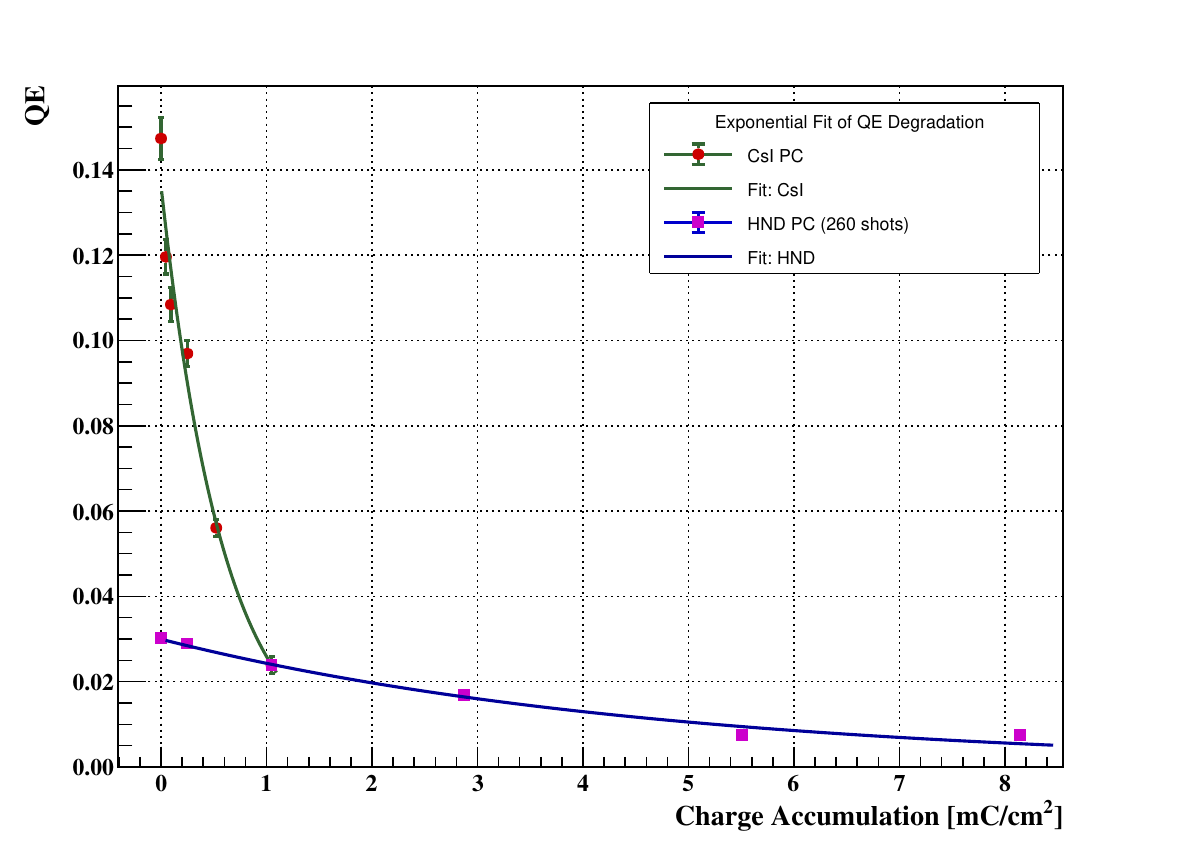}
    \caption{ QE response of CsI and HND photocathodes as a function of charge accumulation in $\sim 10^{-6}$ mbar vacuum.}
    \label{fig:aging}
\end{figure}

\label{sec:hydroganation}
\subsection{ND  Hydrogenation}
HND powder was produced by treating ND grains (average grain size: 250 nm), sourced from D$\&$T (Diamond $\&$ Tools) srl, in a microwave $H_{2}$ plasma using the MWPECVD (Microwave Plasma Enhanced Chemical Vapor Deposition) technique at CNR-ISTP Bari. Details of the hydrogenation process are available in previous publications by Brunbauer et al. \cite{Brunbauer1,brunbauer2,C}. 

\subsection{HND coating}

The coating of ND/HND layers was performed utilizing a pulsed spray technique. The set-up used for this purpose was equipped with an ultrasonic atomizer, a heater and a LabView controlled computer interface. Before coating, HND grains were dispersed in deionized water (in a weight ratio of 1: 1) and sonicated for 30 minutes using a Bandelin Sonoplus HD2070 system. The resulting sonicated solution was then deposited onto PCB substrates (discs of 2.5 cm diameter) using a specified number of spray shots.

During deposition, the substrates were rotated on a magnetic stirrer to ensure an even coating distribution, while the distance between the atomizer nozzle and the substrate surface was fixed at 10 cm. To improve surface coverage and minimize splash effects, the substrates were masked to limit exposure to a 1 cm$^{2}$ area. The PCB discs were heated to 150°C to facilitate rapid evaporation of the solvent ($H_{2}O$) from the surface. Also, for effective $H_{2}O$ evaporation, a 5-second pause was introduced between two successive shots and the duration of the HND spray lasted for 3 seconds. The results presented in this study correspond to coatings applied with 30, 50, and 150 shots and an identical method was used for the coating of THGEMs.

\subsection{aging studies}

Tests of the aging properties of HND and CsI PCs were conducted using the VUV QE measurement setup (ASSET), which was developed by the Gaseous Detectors Development Group (GDD) at CERN~\cite{asset,asset1,asset2}. Before exposure to ions, the QE values of the HND and CsI photocathodes were measured and the samples were then transferred using a linear transfer arm from the QE measurement setup to an irradiation chamber filled with \( Ar/CO_2 \) (70/30\%) gas at ambient pressure. An X-ray beam was introduced into the irradiation chamber to ionize the gas and generate primary electrons. These primary electrons were attracted to multiplication wires, where avalanche multiplication occurred in the presence of an electric field. The avalanche electrons were collected on the wires, while the ions migrated toward the HND and CsI samples. This process led to charge accumulation on the samples, which was monitored alongside the corresponding reduction in QE. 

  Fig. \ref{fig:aging} shows the QE responses of HND and the CsI photocathodes as a function of accumulated charge at fix wavelength of 160 nm. QE of CsI degrades sharply and is reduced by factor ${\sim5}$ at  $1~mC/cm^{2}$ of charge accumulation. The results indicate that HND is approximately an order of magnitude more robust than CsI against ion bombardment. However, as this is a preliminary study, we are unable to pinpoint a specific mechanism responsible for the rapid decline in QE of CsI.

\begin{figure}[h]
    \centering
        \includegraphics[clip, trim=1.0cm 0.3cm 1.5cm 0.0cm, width=0.90\textwidth]{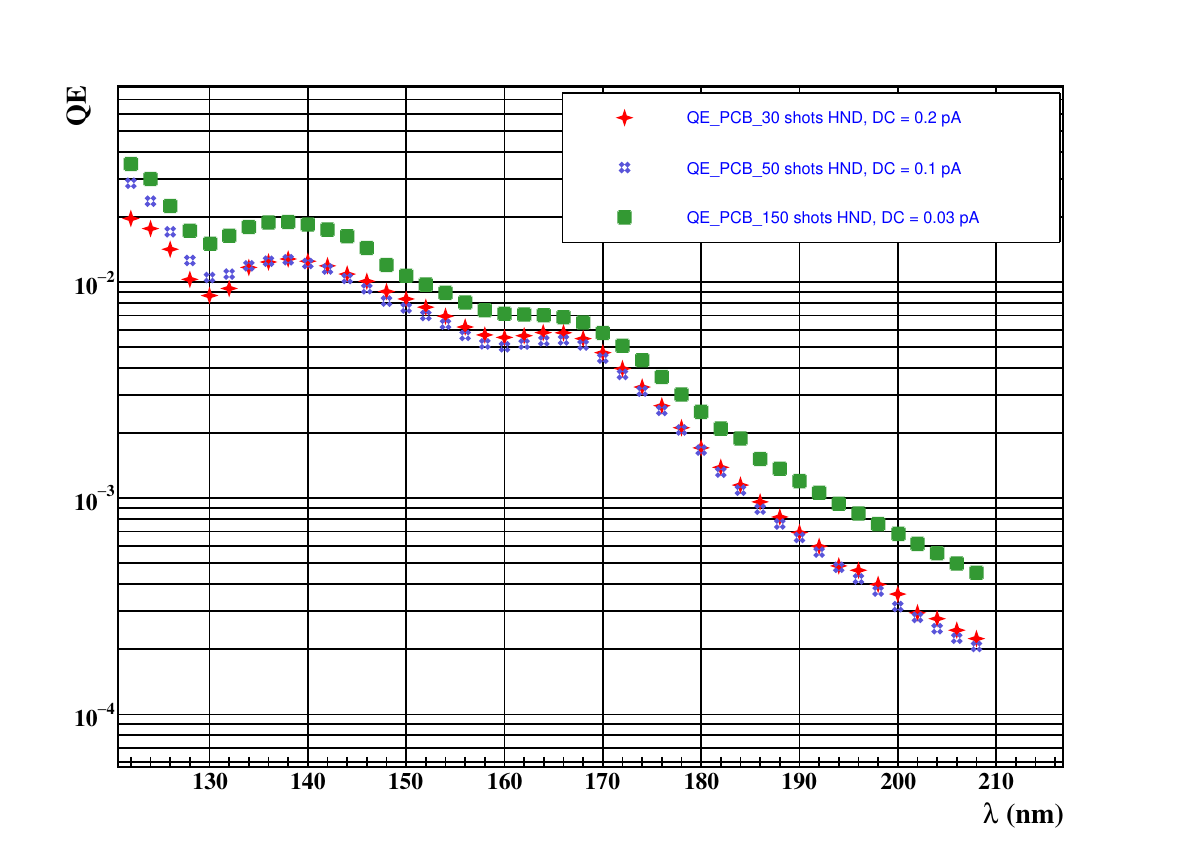}
    \caption{ QE vs. wavelength response of HND photocathodes coated on PCB substrate in vacuum ($\sim 10^{-5} mbar$).}
    \label{Fig:PCBQE}
\end{figure}


\begin{figure}[h]
    \centering
        \includegraphics[clip, trim=0.0cm 0.0cm 0.0cm 0.0cm, width=0.90\textwidth]{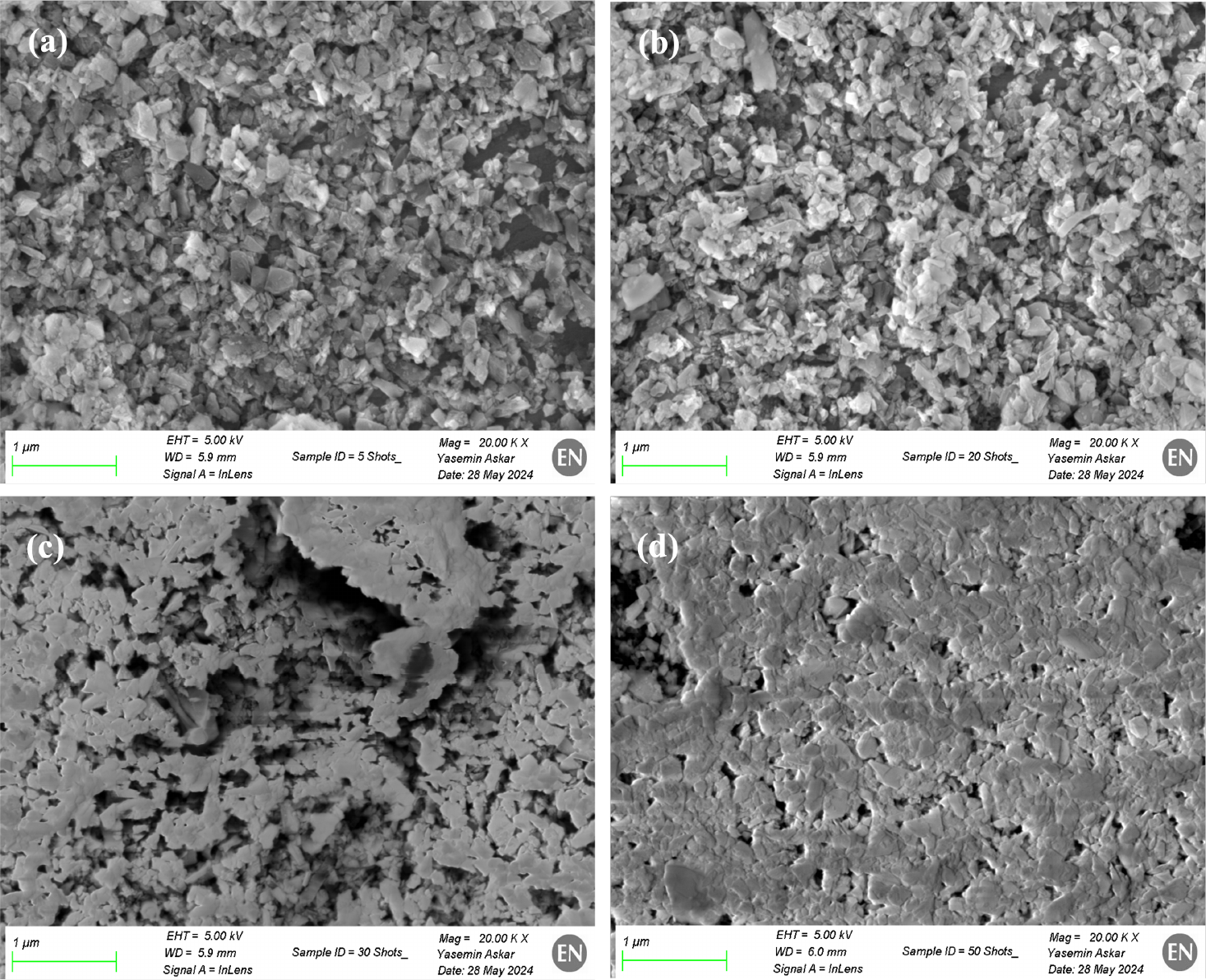}
    \caption{SEM images of HND photocathodes coated on stainless steel substrate with (a) 5, (b) 20, (c) 30 and (d) 50 spray shots.}
    \label{Fig:HNDSEM}
    \end{figure}


\begin{figure}[h]
    \centering
        \includegraphics[clip, trim=0.0cm 0.0cm 0.0cm 0.0cm, width=0.85\textwidth]{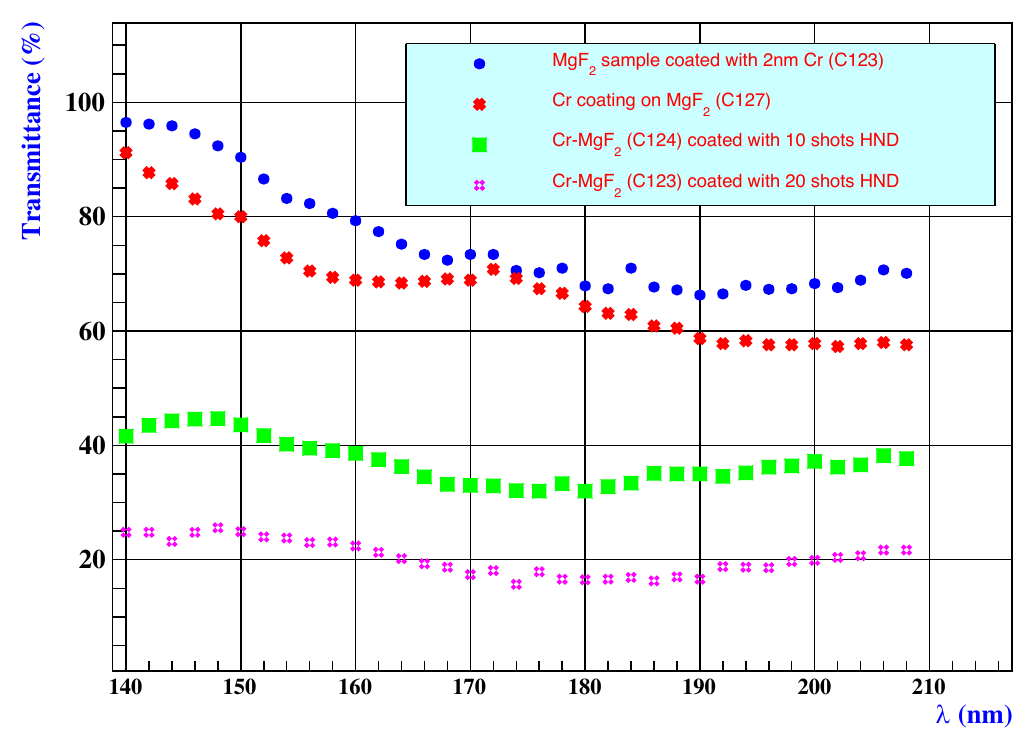}
    \caption{Transmittance spectra of bare and HND coated $Cr-MgF_{2}$ window as a function of wavelength.}
    \label{Fig:trans}
\end{figure}

Several studies in the literature have investigated the aging of VUV-sensitive photocathodes. For example, Singh et al.,  studied ion aging of a 500 nm CsI photocathode, observed a sharp QE decrease from 25\% to about 0.3\% after a charge accumulation of only around 43$\mu$$C/cm^{2}$ at 185 nm wavelength~\cite{Singh}. On the other hand, in their detailed article on photon aging of CsI, J. Va'vra et al. reported a 30\% reduction in QE at same wavelength  from its initial value of 20\% with approximately 30 $\mu$$C/cm^{2}$ charge accumulation~\cite{Jvara}.  Notably, both aging measurements were performed in $CH_{4}$ gas at pressures of 1 atm and 50 bar. However, the evaluation of the long-term performance of CsI photocathodes for the ALICE/HMPID detector by Hoedlmoser et al.\ revealed a reduction in photocurrent at accumulated charge doses of $\geq 1~\mathrm{mC/cm^2}$, whereas doses up to $0.2~\mathrm{mC/cm^2}$ showed no measurable degradation~\cite{Hoed}. These studies further suggest that aging likely results from the interplay of multiple processes occurring both on the surface and within the subsurface layers, which, in turn, alter the electronic structure of the material~\cite{Jvara,Singh}.

\subsection{Photoemission and transmittance measurements}

QE and transmission measurements were carried out at INFN Bari. The set-up included a 30 W $MgF_{2}$ windowed deuterium lamp to emit UV photons of spectral range from 120 to 240 nm and a monochromator (234/302, McPherson) to allow a selection of wavelength for photocurrent measurements. These measurements were performed with the HND-coated PCB disc mounted in a MWPC, under a vacuum of $\sim 10^{-6}$ mbar. A positive voltage of + 90 V was applied to the MWPC wires in order to provide an appropriate electric field at the photoconverter surface  and resulting photocurrent values were measured by a picoammeter (Keithley 617). Before and after the measurements, photocurrent reading from the NIST calibrated silicon photodiode (Model: AXUV-100G) was recorded and used for the estimation of absolute QE.

\begin{figure}[h]
    \centering
        \includegraphics[clip, trim=0.5cm 0.0cm 1.0cm 1.2cm, width=0.8\textwidth]{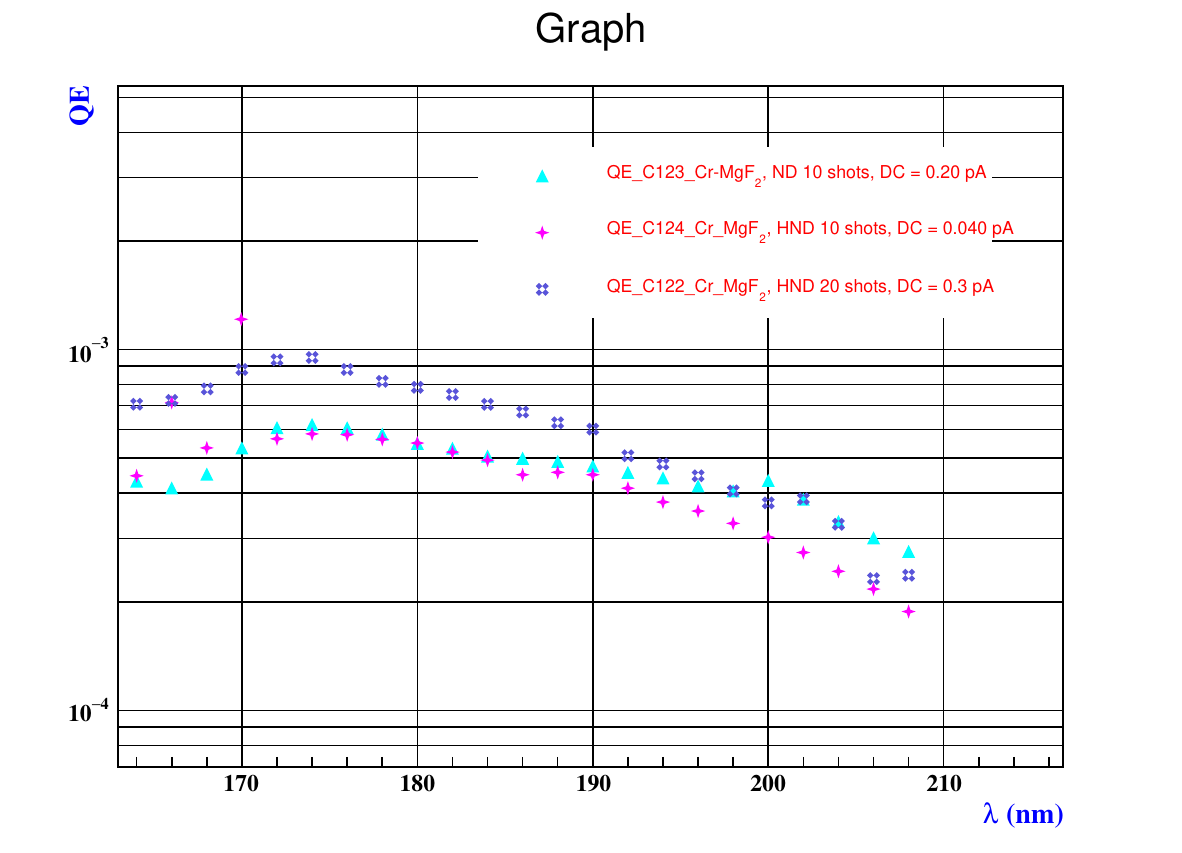}
    \caption{QE vs. wavelength response of Cr-MgF$_{2}$, coated with 10 and 20 spray shots of HND.}
    \label{Fig:QECrMgF2}
\end{figure}
\begin{figure}[h]
    \centering
        \includegraphics[clip, trim=0.0cm 0.0cm 0.0cm 0.0cm, width=0.75\textwidth]{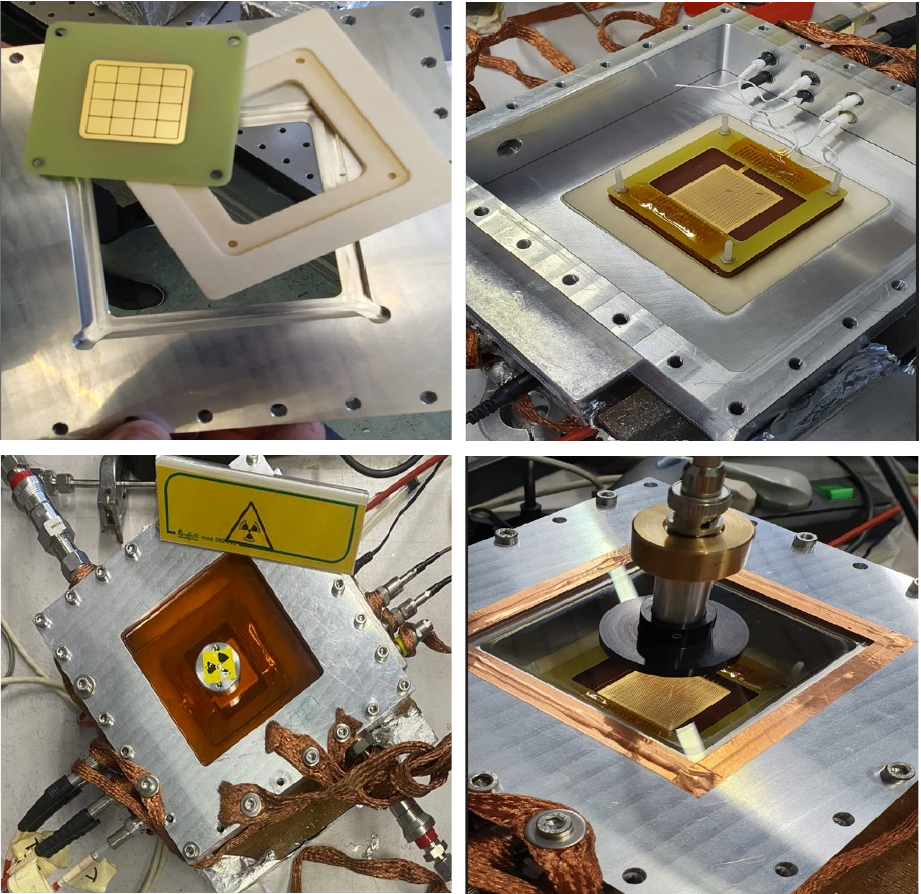}
    \caption{HND based prototype of single photon detector at INFN-Trieste.}
    \label{Fig:prototype}
\end{figure}

\begin{figure}[h]
    \centering
        \includegraphics[clip, trim=0.5cm 0.5cm 0.0cm 0.5cm, width=0.96\textwidth]{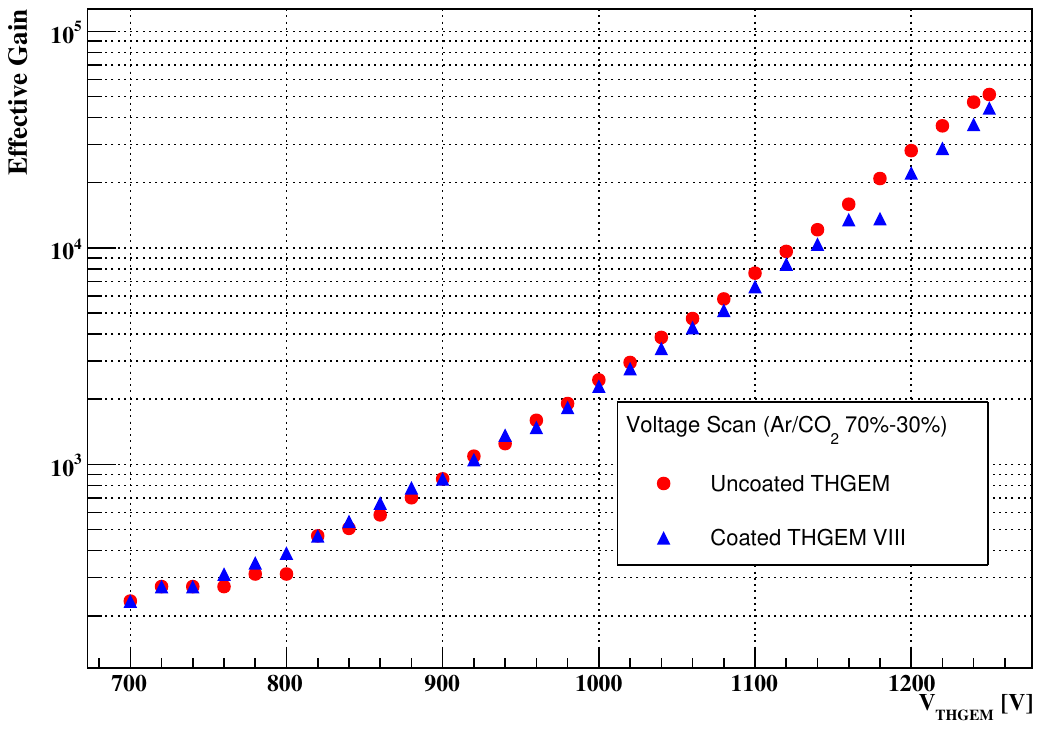}
    \caption{The gain of the prototype detector measured before and after HND coating.}
    \label{Fig:coated}
\end{figure}

The QE measurements of disc samples coated with 30, 50 and 150 shots of HND are presented in Fig. \ref{Fig:PCBQE}. The photocathodes prepared with 30 and 50 shots of HND show similar QE values, lower than those of the sample coated with 150 shots: $\sim4\%$ at 122 nm. The QE value previously obtained for photocathodes prepared with freshly hydrogenated HND grains \cite{Brunbauer1} was a factor $\sim 3$ higher. To investigate the underlying causes of this reduced QE, the same HND solution was coated on stainless steel substrates, and SEM images were obtained at CERN, shown in Fig. \ref{Fig:HNDSEM}. The surface morphology indicates that, although the HND coating is initially non-uniform, the surface area coverage improves with increasing number of shots, leading to a smoother coating surface. Despite these observations, the SEM images do not provide a definitive explanation for the low QE. It is important to note that the HND/H$_{2}$O solution used for the measurements in October 2023 (presented in Fig. \ref{Fig:PCBQE}) underwent hydrogenation and sonication in 2021 and has been stored in water for two years, which may have affected its properties.

To test the feasibility of HND for their application as a photoconverter in future PICOSEC detectors~\cite{bort}, semitransparent HND photocathodes with layers of 10 and 20 shots were deposited on 2 nm $Cr$ coated $MgF_{2}$ window of 1-inch diameter. The window was inserted before the NIST photodiode and current values were recorded. For QE, the obtained current values were normalized to the NIST photodiode response without the window at each wavelength. The initial results on the transmittance and QE values vs wavelength of semitransparent HND photocathodes are shown in Fig. \ref{Fig:trans} and \ref{Fig:QECrMgF2}. In the Fig. \ref{Fig:trans}, the blue  and red curve represent the transmittance of $Cr$ coating, while the green and pink are the transmittance values from $Cr-MgF_{2}$ after coating 10 and 20 shots HND respectively. While, Fig.~\ref{Fig:QECrMgF2} compares the QE values of $Cr-MgF_{2}$ coated with 10 and 20 shots of HND to those of ND powder without hydrogenation.

\subsection{THGEM Characterization}

The characterization of THGEM is carried out at INFN Trieste in a test setup (shown in Fig.~\ref{Fig:prototype}) including a plane of drift wires above the THGEM, a Micromegas electron multiplier and a segmented readout anode plane.  The THGEMs utilized for this work have an active area of $30 \times 30~ mm^{2}$, thickness of $470 ~\mu m$, holes diameter of $0.4 ~mm$, pitch of $0.8~ mm$ and no rim. An $\mathrm{Fe^{55}}$ source was used to generate primaries and the detector signal was read out using a CREMAT CR-110 preamplifier connected to a CREMAT CR-150 r5 evaluation board. The signal was further processed using an Ortec 672 spectroscopy amplifier and digitized using an AMPTEK MCA 8000A, which was also used for data acquisition.

 The effective gain obtained from a single photon prototype detector as a function of THGEM voltage in the mixture of 70\%: 30\% $Ar-CO_{2}$ is shown in Fig. \ref{Fig:coated}. The red points show the gain values prior to HND coating; the blue ones the gain after coating. The characterization of a THGEM, both uncoated and with HND coating, showed nearly identical performance and the maximum stable operational gain values in both cases were about 50k. Therefore, the data reported in Fig. \ref{Fig:coated} confirm that the THGEM response remains unaffected by ND and HND coatings, in agreement with the findings of previous studies \cite{Brunbauer1,brunbauer2}. Systematic studies on HND photocathodes response in different $Ar-CH_{4}$ mixture have also been performed at INFN-Bari and will be reported in detail in a different article.

\section{Conclusion}
In conclusion, the response of HND powder stored in water for two years has been studied, demonstrating significant QE and full compatibility with THGEM operation. 

HND photocathode robustness against ion bombardment was also investigated and found to be 10 times higher than that of CsI.

Semitransparent HND photocathode samples have been produced, yielding encouraging initial results for further testing in picosecond time-resolution configurations.

Further research and development efforts should focus on optimizing HND coating techniques and enhancing QE values through the use of freshly hydrogenated ND powder. 

Additionally, investigating the behavior of HND under various gas mixtures will be essential for unlocking its full potential in practical detector applications.

\section{Acknowledgement}
\begin{itemize}
\item This R\&D activity is partially supported by  EU Horizon 2020 research and innovation programme, STRONG-2020 project, under grant agreement No 824093.
\item  The Program Detector Generic R\&D for an Electron–Ion Collider by Brookhaven National Laboratory, in association with Jefferson Lab and the DOE Office of Nuclear Physics, USA.
\item Author, R. Rai, acknowledges ICTP for providing financial support under TRIL programme. 

\end{itemize}

\end{document}